\documentclass[pdflatex,sn-nature]{sn-jnl}
\usepackage{graphicx}%
\usepackage{multirow}%
\usepackage{amsmath,amssymb,amsfonts}%
\usepackage{amsthm}%
\usepackage{mathrsfs}%
\usepackage[title]{appendix}%
\usepackage{xcolor}%
\usepackage{textcomp}%
\usepackage{manyfoot}%
\usepackage{booktabs}%
\usepackage{lineno}
\usepackage{algorithm}%
\usepackage{algorithmicx}%
\usepackage{algpseudocode}%
\usepackage{listings}%

\usepackage{amssymb}
\usepackage{makecell} 
\usepackage{rotating} 

\theoremstyle{thmstyleone}%
\theoremstyle{thmstyletwo}%

\theoremstyle{thmstylethree}%

\begin{document}

\title[Article Title]{\texttt{QAssemble:} A Pure Python Package for Quantum Many-Body Theory}

\author[1,2]{\fnm{Seongjun} \sur{Mo}}

\author[3]{\fnm{Dongming} \sur{Li}}

\author[2]{\fnm{Mancheon} \sur{Han}}

\author[2,4]{\fnm{Johan} \sur{Jönsson}}

\author[6]{\fnm{Byungkyun} \sur{Kang}}

\author[1]{\fnm{Hoonkyung} \sur{Lee}}

\author[5,7]{\fnm{Gabriel} \sur{Kotliar}}

\author*[2]{\fnm{Sangkook} \sur{Choi}}\email{sangkookchoi@kias.re.kr}

\affil[1]{\orgdiv{Advanced Materials Program, Department of Physics}, \orgname{Konkuk University}, \orgaddress{\city{Seoul}, \postcode{05029}, \country{Korea}}}

\affil[2]{\orgdiv{School of Computational Sciences}, \orgname{Korea Institute for Advanced Study}, \orgaddress{\city{Seoul}, \postcode{02455}, \country{Korea}}}

\affil[3]{\orgdiv{Department of Electrical and Computer Engineering}, \orgname{University of Massachusetts Amherst}, \orgaddress{\street{300 Massachusetts Ave}, \city{Amherst}, \state{MA}, \postcode{01003}, \country{USA}}}

\affil[4]{\orgdiv{Institute of Physics, Faculty of Physics, Astronomy and Informatics}, \orgname{Nicolaus Copernicus University}, \orgaddress{\city{Toruń}, \postcode{87-100}, \country{Poland}}}

\affil[5]{\orgdiv{Condensed Matter Physics and Materials Science Department}, \orgname{Brookhaven National Laboratory}, \orgaddress{Upton, NY} \postcode{11973}, \country{USA}}

\affil[6]{\orgdiv{Department of Physics}, \orgname{The University of Texas at El Paso}, \orgaddress{El Paso, TX} \postcode{79968}, \country{USA}}

\affil[7]{\orgdiv{Department of Physics and Astronomy}, \orgname{Rutgers University}, \orgaddress{NJ}, \postcode{08854}, \country{USA}}

\abstract{QAssemble is a pure-Python package for the quantum many-body 
problem. It implements various functional approaches, such as tight-binding, 
Hartree--Fock, and $GW$ approximations within a unified object-oriented 
architecture. Each physical concept---crystal structure, Hamiltonian, 
Green's function, self-energy, polarizability, screened Coulomb 
interaction---is represented as a distinct class. The modular design 
prioritizes code clarity and extensibility, leveraging NumPy, SciPy and 
libdlr for numerical operations. Performance-critical kernels, including 
the polarizability bubble, Dyson equation inversion, and lattice Fourier 
transforms, are systematically vectorized and combined with the discrete 
Lehmann representation to achieve practical efficiency within a pure-Python 
environment. We validate QAssemble on the electronic structure of graphene 
with local and non-local interactions. Furthermore, benchmarks on a 
five-orbital extended Hund-Hubbard model demonstrate that this strategy 
delivers up to a $60\times$ speedup over traditional loop-based Matsubara 
implementations. QAssemble supports 
both batch execution for production calculations and interactive workflows 
for method development.}


\maketitle

The quantum many-body problem remains one of the central challenges in condensed matter physics. In correlated quantum materials (CQMs), Coulomb repulsion between electrons cannot be treated as a small perturbation; instead, it fundamentally reshapes the electronic structure and gives rise to emergent phenomena\cite{anderson_more_1972,anderson_more_nodate,laughlin_theory_nodate} that lie beyond the reach of single-particle descriptions. These systems exhibit a rich spectrum of collective behaviors—Mott physics\cite{rozenberg_mott-hubbard_1992,zhang_mott_1993}, Hund-metal physics\cite{de_medici_janus-faced_2011,stadler_differentiating_2021,ryee_hund_2021}, heavy-fermion physics\cite{jang_evolution_2020}, and unconventional superconductivity\cite{mathur_magnetically_1998,jiao_chiral_2020}, among others. Accurately capturing electron correlations is therefore essential for understanding the physics of these materials.

The functional approach offers a powerful framework for tackling this challenge, complementing wavefunction-based methods\cite{kent_toward_2018}. Rather than solving for the full many-body wavefunction—whose complexity grows exponentially with system size—functional techniques work directly with Green's functions. These methods rest on a free-energy functional formulation in which the exact solution is expressed as a functional of the Green's function and self-energy. Crucially, Green's functions provide a direct link to experimental 
observables: photoemission spectroscopy probes the one-particle Green's 
function, while optical and other response functions involve two-particle 
correlators. This functional formalism originated with Luttinger and Ward~\cite{luttinger_ground-state_1960} and Baym and Kadanoff~\cite{baym_conservation_1961}, and was later extended~\cite{almbladh_variational_1999} to incorporate screened Coulomb interactions and irreducible polarizabilities.

Within this framework, a hierarchy of diagrammatic approximations provides systematic control over electron correlations. Hartree–Fock (HF)~\cite{hartree_self-consistent_1997,slater_self_1928,slater_note_1930,slater_simplification_1951,froese_fischer_general_1987} captures fermionic exchange at the mean-field level. The $GW$ approximation~\cite{hedin_new_1965,hybertsen_electron_1986,godby_self-energy_1988,massidda_band-structure_1995,massidda_quasiparticle_1997,aulbur_quasiparticle_2000,fulde_semiconductors_1995} goes beyond HF by including dynamical screening of the Coulomb interaction. Dynamical mean-field theory (DMFT)~\cite{kotliar_strongly_2004,georges_dynamical_1996,kotliar_electronic_2006,anisimov_electronic_2010,imada_electronic_2010} treats local quantum fluctuations exactly. $GW$+EDMFT methods~\cite{sun_extended_2002,biermann_first-principles_2003,kang_comdmft_2025} combine the strengths of $GW$ and DMFT by treating local and nonlocal correlations on an equal footing.

Despite their theoretical elegance, implementing these many-body methods poses significant practical challenges. The algorithms involve nested summations, frequency–time transforms, and self-consistency loops that are difficult to code correctly. Validation is equally demanding: analytical solutions exist only for toy models, and comparisons between codes often reveal discrepancies arising from implementation details or differing conventions. Creating user-friendly interfaces that hide this complexity without sacrificing flexibility remains a persistent obstacle. Several software packages address aspects of these issues: PythTB~\cite{yusufaly_tight-binding_nodate} provides Python tools for tight-binding calculations, ALPS~\cite{alet_alps_2005} offers a library for quantum lattice simulations, and TRIQS~\cite{parcollet_triqs_2015} supplies powerful infrastructure for diagrammatic many-body methods.

In this paper we present QAssemble, a Python framework designed to make many-body calculations more accessible while incorporating state-of-the-art algorithms. Built on object-oriented principles, QAssemble emphasizes code clarity and modularity: each physical concept—crystal structure, Hamiltonian, Green's function, self-energy—is represented as a distinct class with well-defined interfaces. This design prioritizes readability and extensibility over raw performance, though performance-critical operations leverage optimized libraries such as NumPy~\cite{harris2020array}, SciPy~\cite{2020SciPy-NMeth}, and libdlr~\cite{kaye_libdlr_2022}. The framework has been validated through benchmarks against existing codes and established numerical results, and supports both batch processing for production runs and interactive Jupyter notebooks for exploration and teaching.

A distinguishing design choice of QAssemble is its commitment to a fully pure-Python implementation. Established many-body frameworks typically delegate performance-critical routines to compiled Fortran or C++ kernels wrapped by a Python interface. Such designs deliver excellent performance, but the computational core remains effectively opaque to users who are not fluent in the underlying compiled language or its build system. QAssemble explores a complementary route: the entire codebase, from lattice Fourier transforms and Dyson equation inversion to the polarizability bubble and the construction of the screened interaction, is written in Python, with performance recovered through systematic vectorization on top of NumPy, SciPy, and libdlr. This choice is motivated by two considerations. First, transparency: every step of a diagrammatic calculation is directly inspectable in a single language. This lowers the barrier for verification, pedagogy, and cross-code comparison. Furthermore, it creates an immediate correspondence between the formal equations in the Methods section and the actual numerical kernels. Second, hackability: researchers can prototype new self-energy diagrams, alternative self-consistency schemes, or extensions by modifying Python classes directly, without recompiling intermediate libraries or managing a foreign build environment. We regard this as a deliberate trade-off in favor of a collaborative, community-driven development model, and we demonstrate in the Performance benchmark section that vectorization combined with the discrete Lehmann representation suffices to keep the resulting framework practically competitive, achieving up to roughly 60$\times$ speedup over a loop-based Matsubara implementation on a representative five-orbital benchmark.

At present, QAssemble implements tight-binding band-structure calculations together with self-consistent Hartree–Fock and $GW$ approximations. Its modular architecture is designed to accommodate future extensions, including DMFT for strong local correlations, EDMFT, $GW$+EDMFT, and beyond. By providing an accessible platform built on modern algorithms, QAssemble aims to lower the barrier to entry for many-body electronic-structure calculations while retaining the sophistication required for research applications.

\section{Results}\label{sec2}

QAssemble consists of seven core classes implemented as Python modules: CorrelationFunction, Crystal, DLR, FLatDyn, FLatStc, BLatDyn, and BLatStc. These components are controlled by the driver script "QAssemble.py".

\begin{figure*}
  \centering
  \includegraphics[width=0.75\columnwidth]{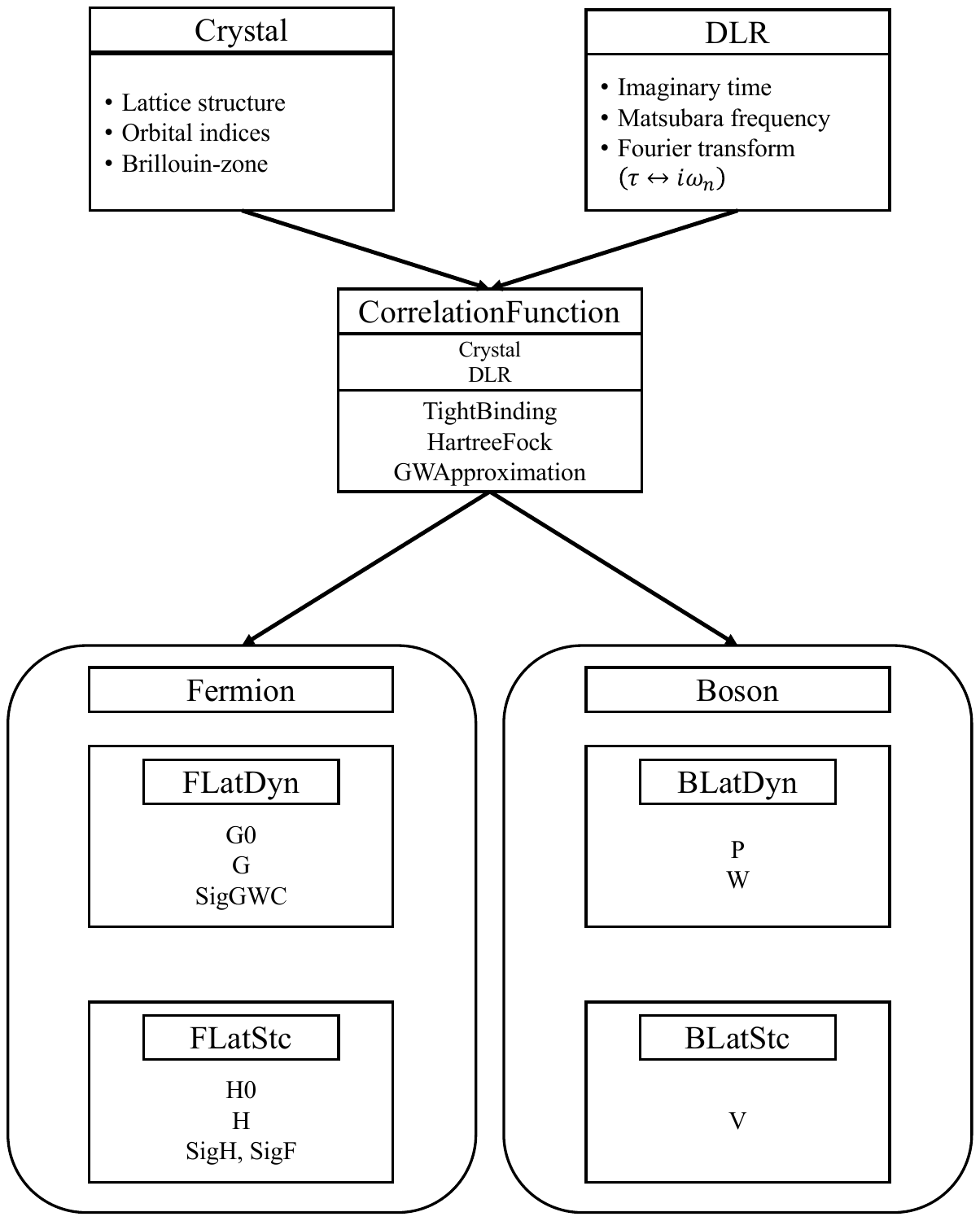}

  \caption{\textbf{QAssemble architecture and workflow for many-body electronic structure calculations.} Crystal and DLR components define lattice/orbital structures and imaginary-time/Matsubara grids. CorrelationFunction orchestrates tight-binding (TB), HF, and $GW$ methods via fermionic (FLatStc, FLatDyn) and bosonic (BLatStc, BLatDyn) modules}
  \label{fig-classdiagram}
\end{figure*}

Figure \ref{fig-classdiagram} illustrates the QAssemble architecture. Adopting a pure-Python object-oriented paradigm, the design emphasizes clarity, modularity, and user accessibility. The architecture distinguishes between foundational data structures—such as crystal information (Crystal) and imaginary time-frequency sampling (DLR)—and physical quantities. These physical objects are organized along two orthogonal axes: temporal dependence (static vs. dynamic) and particle statistics (fermionic vs. bosonic). This structure yields four distinct object categories that encapsulate specific many-body calculations, including Hamiltonians, Green's functions, self-energies, and polarizabilities, each with explicit methods for computation, transformation, and persistence.

\subsection{Core class}

\subsubsection{FLatDyn class}
FLatDyn is the base class for frequency-dependent fermionic quantities on the lattice. Constructed with both Crystal and DLR instances, it stores data in DLR representation. It provides common functions for dynamic fermionic quantities, including Fourier transforms between $\mathbf{k}$-space and real space, Fourier transforms between Matsubara frequency and imaginary time, solving the Dyson equation, inverse calculations at a given crystal momentum and matsubara frequency. Figure \ref{fig-flatdyndiagram} (a) summarizes the structure of FLatDyn class and its child classes such as G$0$, G, and SigGWC. 

\subsubsection{FLatStc class}
FLatStc is the base class for frequency-independent fermionic quantities on the lattice. Constructed with Crystal instances. It provides common functions for static fermionic quantities, including Fourier transforms between $\mathbf{k}$-space and real space, inverse calculations, and diagonalization calculations. Figure \ref{fig-flatdyndiagram} (b) summarizes the structure of FLatDyn class and its child classes including H0, H, SigH, and SigF.

\subsubsection{BLatDyn class}
BLatDyn is the base class for frequency-dependent bosonic quantities on the lattice. Constructed with both Crystal and DLR instances, it stores data in DLR representation. It provides common functions for dynamic bosonic quantities, including Fourier transforms between $\mathbf{k}$-space and real space, Fourier transforms between Matsubara frequency and imaginary time, solving the Dyson equation, inverse calculations at a given crystal momentum and matsubara frequency. Figure \ref{fig-flatdyndiagram} (c) summarizes the structure of BLatDyn class and its child classes such as P, and W.

\subsubsection{BLatStc class}
\begin{figure*}
  \centering
    \includegraphics[width=1\columnwidth]{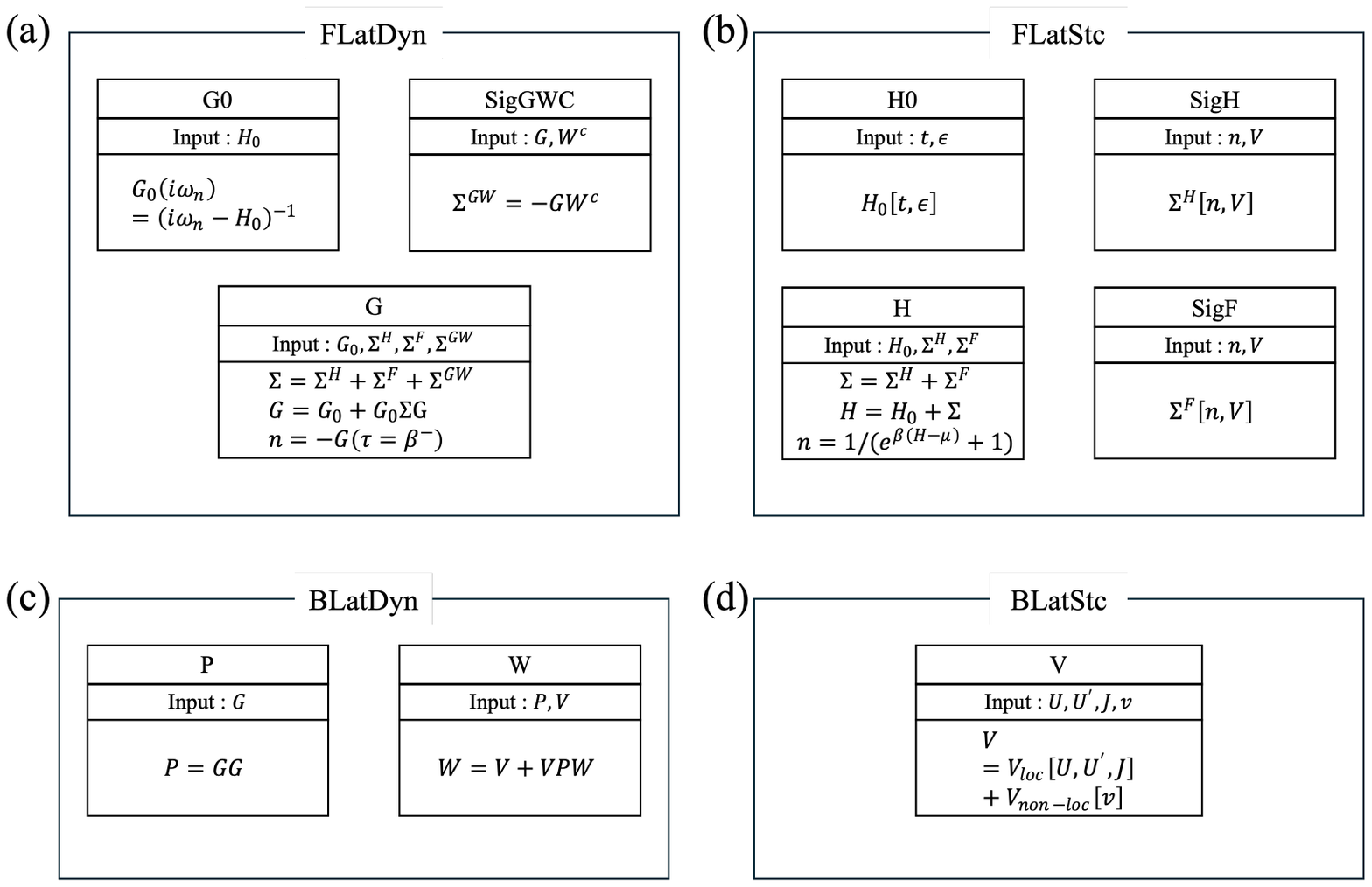}
  \caption{\textbf{Core class hierarchy.} Class diagram for (a) FLatDyn, (b) FLatStc, (c) BLatDyn, and (d) BLatStc}
  \label{fig-flatdyndiagram}
\end{figure*}

BLatStc is the base class for frequency-independent bosonic quantities on the lattice. Constructed with Crystal instances. It provides common functions for static bosonic quantities, including Fourier transforms between $\mathbf{k}$-space and real space, inverse calculations, and diagonalization calculations. Figure \ref{fig-flatdyndiagram} (d) summarizes the structure of BLatDyn class and its child class, V.

\subsection{Self-consistent iteration schemes}

The classes described above serve as building blocks for self-consistent many-body calculations. Both HF and $GW$ approximations require iterative solutions. This is because the self-energy $\Sigma$ depends on the Green's function $G$, while the Green's function is determined by $\Sigma$ via the Dyson equation. We now describe how these classes orchestrate the self-consistent loops.

\begin{figure*}[t]
  \centering
  \includegraphics[width=0.72\columnwidth]{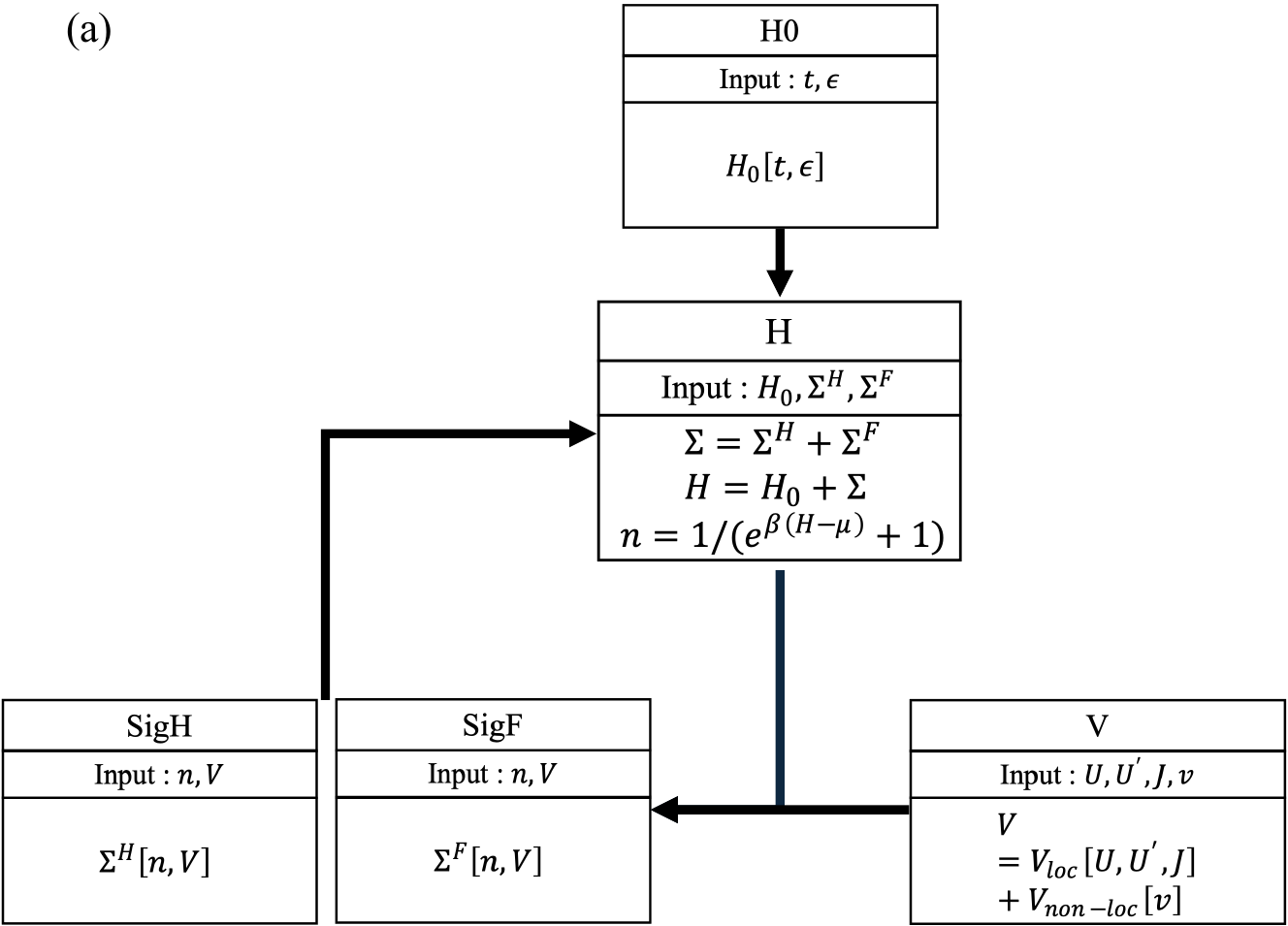}\par\vspace{2mm}
  \includegraphics[width=0.72\columnwidth]{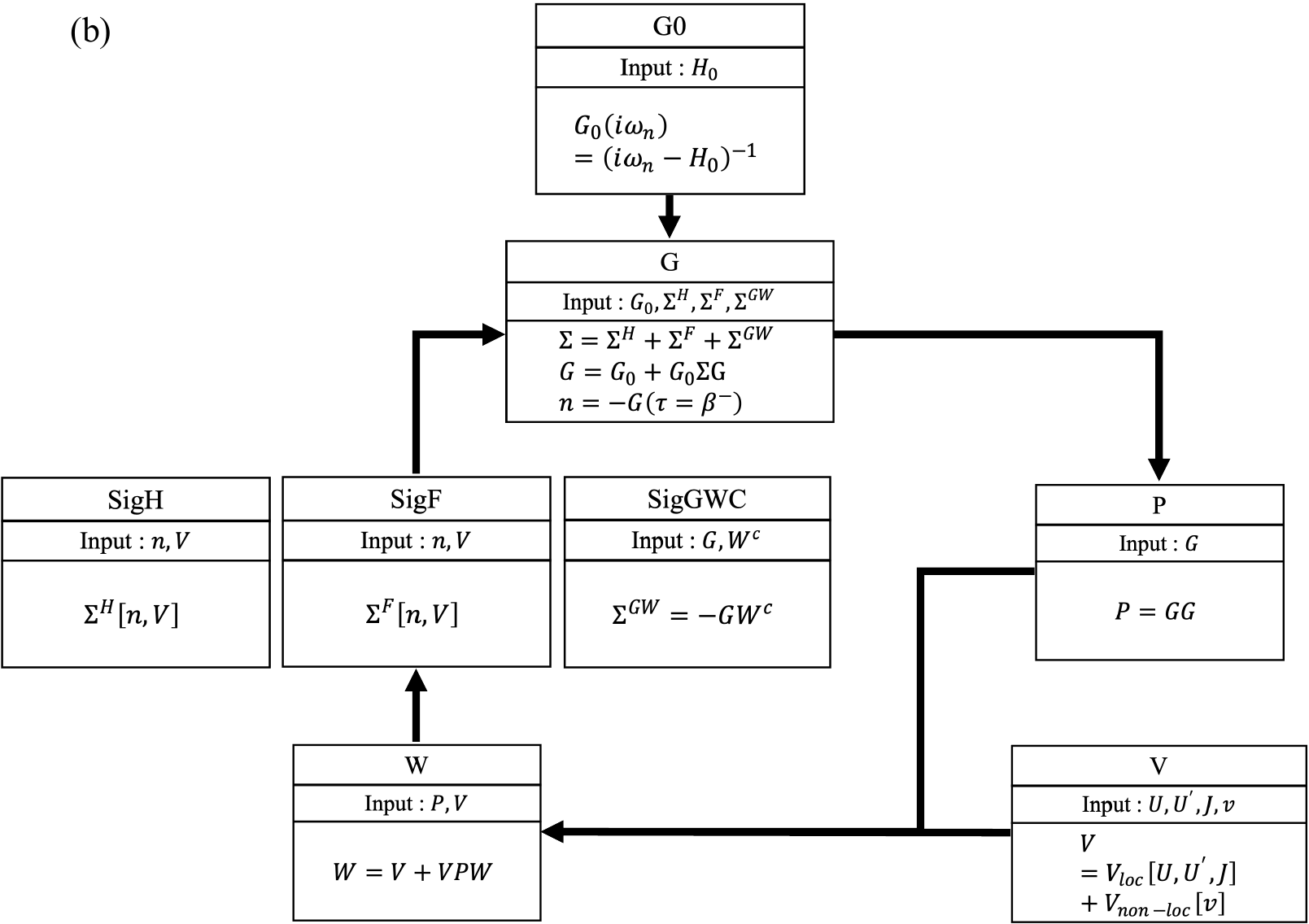}
  \caption{\textbf{Self-consistent workflows for Hartree–Fock and $GW$ approximations.} (a) The HF loop iterates between H0, V, SigH, SigF, and Hamiltonian classes to solve $H = H_0 + \Sigma^H + \Sigma^F$ self-consistently for the density matrix $n$. (b) The $GW$ loop extends (a) by additionally computing the irreducible polarizability $P = GG$ (P), the dynamically screened interaction $W = V + VPW$ (W), and the correlation self-energy $\Sigma^{GW} = -GW^C$ (SigGWC), with G combining all contributions via the Dyson equation. Boxes indicate classes; arrows denote data flow between components.}
  \label{fig-self}
\end{figure*}

\subsubsection{Hartree-Fock self-consistency}

Figure \ref{fig-self} (a) illustrates the Hartree-Fock self-consistent workflow involving five classes: H0, H, SigH, SigF, and V. The Hamiltonian class acts as the central hub, aggregating self-energy contributions to construct the effective single-particle Hamiltonian at each iteration. The workflow proceeds through the following components.

\paragraph{H0 class}

The H0 class constructs the non-interacting Hamiltonian $H_0$ from the lattice geometry (provided by Crystal), hopping amplitudes $t$, and on-site energies $\epsilon$. It returns $H_0[t,\epsilon]$ as defined in Eq. \ref{eq:many-body-hamiltonian}, stored as a frequency-independent FLatStc object that remains fixed throughout the self-consistent cycle.

\paragraph{V class}

The V class constructs the bare Coulomb interaction $V$ from user-specified parameters. Local interactions are handled through Slater\cite{slater_quantum_1960} or Kanamori\cite{kanamori_electron_1963} parametrizations, with support for transformations between the two\cite{van_der_marel_electron-electron_1988,strand_valence-skipping_2014}. Non-local interactions are specified either explicitly through density-to-density couplings $V_{ij}$ or generated from model potentials such as the Ohno form\cite{ohno_remarks_1964}. The class returns $V$ as a BLatStc object.

\paragraph{SigH class}

The SigH class computes the Hartree self-energy (Eq. \ref{eq-hartree-self-energy}). This contribution captures the direct Coulomb repulsion in a mean-field approximation. The class accepts the density matrix $n$ and bare interaction $V$ as inputs and returns $\Sigma^H$ as an FLatStc object.

\paragraph{SigF class}

The SigF class computes the Fock self-energy (Eq. \ref{eq-fock-self-energy}), which is generally off-diagonal in both spatial and orbital indices. This term accounts for exchange contributions arising from the antisymmetry of the many-electron wavefunction. Like SigH, it accepts $n$ and $V$ as inputs and returns $\Sigma^F$ as an FLatStc object. Together, $\Sigma^H$ and $\Sigma^F$ feed back into the Hamiltonian class for the next iteration, continuing until convergence is achieved.

\paragraph{H class}

At each step $i$, it receives $H_0$ (fixed), the Hartree self-energy $\Sigma^H$ from SigH, and the Fock self-energy $\Sigma^F$ from SigF. Then it combines them into $H = H_0 + \Sigma^H + \Sigma^F$. The chemical potential $\mu$ is then adjusted to satisfy the target electron filling, and the density matrix $n$ is computed via the Fermi-Dirac distribution $n = 1/(e^{\beta(H - \mu)}+1)$. The class returns the updated Hamiltonian $H$, density matrix $n$, and chemical potential $\mu$.

\paragraph{$GW$ self-consistency}

Figure \ref{fig-self} (b) illustrates the GW self-consistent workflow, which extends the Hartree-Fock calculation (Figure \ref{fig-self} (a)) by incorporating dynamically screened Coulomb interactions. The GW loop introduces three additional classes—P, W, and SigGWC—that work in concert with the existing G, SigH, and SigF classes to capture correlation effects beyond static mean-field theory. The key conceptual advance is that the static interaction $V$ is replaced by a frequency-dependent screened interaction $W$ that accounts for the polarization response of the electron gas. We now describe how each component contributes to this extended workflow.

\paragraph{G0 class}

The calculation begins with the non-interacting Green's function $G_0$ (Eqs. \ref{eq:bare-green-tau} and \ref{eq:bare-green-freq}), constructed by G0 from the non-interacting Hamiltonian $H_0$ produced earlier by NIHamiltonian (Figure \ref{fig-self} (a)). This establishes the starting point for the self-consistent iteration.

\paragraph{P class}

P computes the irreducible polarizability $P$ using Green's function by evaluating the two-particle correlation function (Eq. \ref{eq-polarizability}). This convolution of two fermionic Green's functions produces the bosonic response function that describes how the electron density responds to screened perturbations.

\paragraph{W class}

W then constructs the dynamically screened Coulomb interaction $W$ from $P$ and the bare interaction $V$ via the Dyson equation (Eq. \ref{eq-screened-coulomb}). Formally, it represents a resummation of bubble diagrams where the interaction line is repeatedly dressed by polarization insertions, describing how the bare Coulomb potential is renormalized by the surrounding electron gas.

\paragraph{SigGWC class}

SigGWC computes the GW correlation self-energy (Eq. \ref{eq-gw-self-energy}) by convolving the Green's function $G$ with the correlation part of the screened interaction $W^C = W - V$. The subtraction isolates the dynamical screening contribution, avoiding double-counting since static Coulomb interaction effects are already incorporated through $\Sigma^F$.

\paragraph{G class}

At each iteration, it receives $G_0$ from G0 along with three self-energy contributions: the Hartree term $\Sigma^H$ from SigH, the Fock term $\Sigma^F$ from SigF, and the GW correlation self-energy $\Sigma^{GWC}$ from SigGWC. These are combined via the Dyson equation (Eq. \ref{full-green-freq}) to yield the fully interacting Green's function $G$. The chemical potential $\mu$ is then adjusted to maintain the target filling, and the density matrix is extracted from $n = -G(\tau=\beta^{-})$. G returns the updated $G$, $n$, and $\mu$, with the new $G$ feeding back into P to begin the next iteration.

\paragraph{Z class}

Z extracts the quasiparticle renormalization factor from the converged GW self-energy. It receives the self-energy $\Sigma(\mathbf{k}, i\omega_n)$ stored in DLR representation and computes the inverse renormalization factor via
\begin{equation}
  Z(\mathbf{k})^{-1} = 1 - \left. \frac{\partial \Sigma(\mathbf{k}, i\omega_n)}{\partial i\omega_n} \right|_{\omega=0}.
  \label{eq-z-factor}
\end{equation}
Z returns the $\mathbf{k}$-resolved renormalization factor $Z(\mathbf{k})$, which encodes the dynamical mass enhancement and spectral weight transfer from coherent quasiparticle peaks to incoherent satellite structures.

\paragraph{SigStc class}

SigStc computes the static limit of the self-energy $\Sigma(\mathbf{k}, \omega=0)$ from the full frequency-dependent GW result. It receives the self-energy $\Sigma(\mathbf{k}, i\omega_n)$ in DLR representation and evaluates its zero-frequency limit. This static self-energy captures the shifts in quasiparticle energies. SigStc returns the $\mathbf{k}$-resolved static self-energy matrix, which serves as an input for constructing the quasi-particle Hamiltonian.


\subsection{Implementation detail}
\subsubsection{Vectorized numerical kernels}\label{sec:vectorization}

QAssemble systematically vectorizes the evaluation of physical quantities---including phase factors in lattice Fourier transforms, the polarizability Eq. \ref{eq-polarizability}, basis index mappings, and the solving Dyson equation Eq. \ref{full-green-freq}---by expressing them as batched array operations over momentum, frequency, and orbital indices using NumPy. This approach eliminates explicit Python loops over these indices, expressing the computational work as bulk array operations through NumPy's vectorized interface.
As a result, practical efficiency is achieved within a pure-Python environment without sacrificing code clarity. Quantitative performance benchmarks are presented in Section~\ref{sec:validation}.




\subsubsection{Installation and Execution Modes}

QAssemble is distributed as a standard Python package built with a 
\texttt{pyproject.toml} specification, and is installed by cloning the 
source repository and running \texttt{pip install -e .} from the project 
root. The editable installation is recommended because it allows users to 
modify class implementations or prototype new self-energy diagrams without 
reinstalling the package, in line with the hackability that motivates 
the pure-Python design. The installation pulls in the standard scientific 
Python stack (NumPy, SciPy, h5py, mpi4py, libdlr) and exposes the 
\texttt{qassemble} command-line entry point.

Once installed, QAssemble offers two complementary execution modes. 
\textit{Batch mode} is designed for script-based execution with all 
parameters specified in a single input file (\texttt{input.ini}), and is 
launched from the command line as
\begin{verbatim}
$ qassemble
\end{verbatim}
This mode is suited to automated production calculations on HPC clusters: 
all output is written to HDF5 with the input file embedded in the file 
metadata, ensuring full reproducibility of every calculation. 
\textit{Interactive mode} exposes the same class hierarchy directly to a 
Python or IPython session, allowing step-by-step construction of the 
Hamiltonian, Green's function, and self-energy objects with real-time 
inspection of intermediate quantities. This mode is particularly natural 
inside Jupyter notebooks, where it supports method development and 
on-the-fly diagnosis of convergence behavior.




\subsection{Validation: Electronic structure of graphene}
\label{sec:validation}

We validate QAssemble by computing the electronic structure of graphene at temperature $T = 2000\ \text{K}$ using three levels of theory: tight-binding (TB), Hartree-Fock (HF), and $GW$. The $GW$ calculation employs the discrete Lehmann representation with cutoff energy $\Lambda = 100\ \text{eV}$ and a $25 \times 25 \times 1$ $k$-point grid. The system is described by the extended Hubbard model:
\begin{equation}
    \begin{split}
        H = &-t \sum_{\langle ij \rangle \sigma} \hat{c}^{\dagger}_{i\sigma} \hat{c}_{j\sigma} + \frac{U}{2} \sum_{i} \sum_{\sigma \sigma'} \hat{c}^{\dagger}_{i\sigma} \hat{c}^{\dagger}_{i\sigma'} \hat{c}_{i\sigma'} \hat{c}_{i\sigma} \\
        &+ \frac{V}{2} \sum_{\langle ij \rangle} \sum_{\sigma \sigma'} \hat{c}^{\dagger}_{i\sigma} \hat{c}^{\dagger}_{j\sigma'} \hat{c}_{j\sigma'} \hat{c}_{i\sigma},
    \end{split}
    \label{eq-extended-hubbard}
\end{equation}
where $t = 1.0\ \text{eV}$ is the nearest-neighbor hopping amplitude, $U = 2.0\ \text{eV}$ is the on-site Coulomb repulsion, and $V = 0.2\ \text{eV}$ is the nearest-neighbor Coulomb interaction. The system is maintained at half-filling throughout all calculations.

\subsubsection{Electronic structure within TB, HF, and GW}

\begin{figure}
    \includegraphics[width=1.\textwidth]{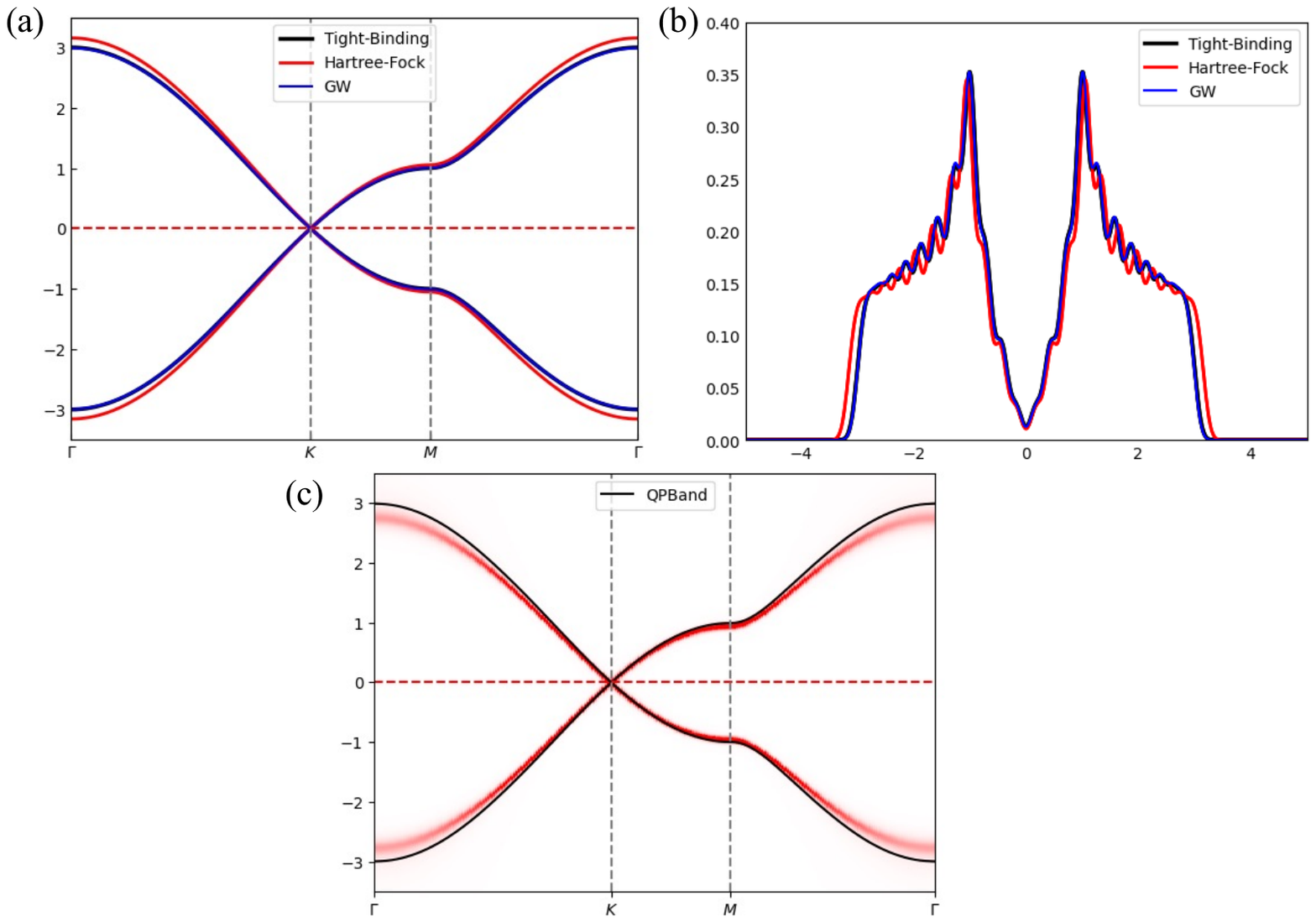}
    \caption{\textbf{Electronic structure of graphene.} (a) Band structure along high-symmetry paths comparing TB, HF, and $GW$ results. (b) Density of states obtained by integration over a $20 \times 20 \times 1$ Brillouin zone mesh. (c) $A(\mathbf{k}, \omega)$ at $T = 2000\ \text{K}$ along the high-symmetry path $\Gamma$-K-M-$\Gamma$. The color scale indicates spectral function, with the overlaid black curve showing the quasiparticle band structure obtained from diagonalization of $H_{\mathrm{QP}}(\mathbf{k})$ in eq. \eqref{eq-quasi-particle-Hamiltonian}}
    \label{fig-band-dos}
\end{figure}

Figure~\ref{fig-band-dos}(a) displays the evolution of the graphene band structure from the non-interacting tight-binding limit through increasingly sophisticated treatments of electron-electron interaction. The TB calculation reproduces the familiar Dirac-cone dispersion characteristic of graphene, with the valence and conduction bands meeting at the $K$ and $K'$ points of the Brillouin zone to form a zero-gap semimetal. The linear dispersion near these points gives rise to the well-known massless Dirac fermion behavior.

HF level introduces quantitative modifications while preserving the qualitative semimetallic character. The Hartree term produces an energy shift corresponding to the classical electrostatic potential, while the Fock exchange introduces momentum-dependent corrections that slightly modify the bandwidth. At the chosen interaction strength, these corrections remain modest—the Dirac points stay protected by symmetry and the system remains metallic.

The full $GW$ calculation captures dynamical screening effects absent in the static HF treatment. The frequency-dependent screened interaction $W$ modifies the self-energy beyond the instantaneous Hartree and Fock contributions, producing a further bandwidth reduction relative to HF consistent with enhanced screening of the bare Coulomb interaction. To represent these dynamical effects within a static framework, we construct the quasi-particle Hamiltonian

\begin{equation}
  \begin{split}
    H_{\mathrm{QP}}(\mathbf{k}) = \sqrt{Z(\mathbf{k})} \left( H_0(\mathbf{k}) + \Sigma(\mathbf{k}, \omega=0) \right) \sqrt{Z(\mathbf{k})},
  \end{split}
  \label{eq-quasi-particle-Hamiltonian}
\end{equation}
where $H_0(\mathbf{k})$ is the non-interacting tight-binding Hamiltonian, $\Sigma(\mathbf{k}, \omega = 0)$ is the static self-energy obtained from SigmaStc, and $Z(\mathbf{k})$ is the renormalization factor computed by Zfactor. The symmetric placement of $\sqrt{Z(\mathbf{k})}$ ensures Hermiticity when $Z(\mathbf{k})$ is a matrix in orbital space. Diagonalizing $H_{\mathrm{QP}}(\mathbf{k})$ yields the renormalized quasiparticle band structure shown in Fig.~\ref{fig-band-dos}(a), incorporating both the static self-energy shift and the dynamical mass enhancement encoded in $Z(\mathbf{k})$.

\subsubsection{Density of states}

Figure~\ref{fig-band-dos}(b) presents the density of states computed within the three levels of theory, obtained by integrating the spectral function over a $20 \times 20 \times 1$ Brillouin zone mesh. This uniform $k$-space sampling reveals how electron correlation reshapes the electronic structure across the entire zone, rather than only along high-symmetry paths.

The TB DOS exhibits the characteristic semi-metallic behavior of graphene, with a V-shaped dip approaching zero at the Fermi level and symmetric van Hove singularities at $\epsilon \approx \pm 1\ \text{eV}$ arising from saddle points at the M points. The HF DOS shows moderate quantitative modifications: the van Hove singularities shift slightly in energy due to the momentum-dependent Fock exchange, while the V-shape dip at the Fermi level remains pronounced. The overall shape is preserved, indicating that static mean-field correlations do not fundamentally alter the electronic structure in this parameter regime. The $GW$ DOS exhibits band with narrowing from the HF results. Dynamical screening supresses the effect of Fock-terms. The DOS profile becomes much like TB results.

\subsubsection{Spectral function}


To obtain the real-frequency electronic structure, we compute the $\mathbf{k}$-resolved spectral function $A(\mathbf{k}, \omega)$ from the converged $GW$ Green's function via analytic continuation\cite{PhysRevB.92.060509}. The Matsubara-axis data is continued to real frequencies using the parameter-free maximum entropy method\cite{PhysRevB.106.245150}.

Figure~\ref{fig-band-dos} (c) shows the spectral function along the high-symmetry path overlaid with the quasiparticle bands (red curves) obtained from diagonalizing $H_{\mathrm{QP}}(\mathbf{k})$. Sharp quasiparticle peaks trace the renormalized dispersion, with spectral weight concentrated along well-defined bands throughout the Brillouin zone. At the Dirac points K and K$'$, the spectral function exhibits the characteristic linear dispersion with well-defined peaks crossing the Fermi level (dashed horizontal line at $\omega = 0$), confirming that the $GW$ approximation preserves the massless Dirac fermion character of low-energy excitations in graphene. Away from the Dirac points, particularly near the M point where van Hove singularities are expected, the peaks remain sharp and the spectral weight remains strongly peaked near the quasiparticle dispersion. The bandwidth and overall dispersion reflect the self-energy corrections characteristic of the $GW$ approximation, which accounts for dynamical screening effects beyond mean-field theory.

\subsection{Performance benchmark: Degenerate five-orbital extended Hund-Hubbard model}

To assess the performance of QAssemble beyond the minimal two-orbital graphene case, we benchmark the four computational configurations on a degenerate five-orbital extended Hund-Hubbard model with nearest-neighbor hopping $t = 1.0$\,eV, on-site Coulomb repulsion $U = 1.0$\,eV, Hund's coupling $J = 0.1$\,eV, and nearest-neighbor Coulomb interaction $V = 0.2$\,eV, computed on an $8\times8\times8$ $\mathbf{k}$-grid with inverse temperature $\beta = 100$\,eV$^{-1}$ and a Matsubara cutoff of $\Lambda = 25$\,eV.
This system is representative of correlated $d$-electron materials where the orbital dimension introduces a significantly heavier computational burden.

\begin{table}[h]
\centering
\caption{Wall-clock time (seconds) per GW iteration for each physical quantity, measured on the five-orbital extended Hund-Hubbard model ($8\times8\times8$ $k$-grid, $\Lambda=25$\,eV, $\beta=100$\,eV$^{-1}$). MF denotes the conventional Matsubara frequency grid with Fortran-based core routines; DLR denotes the discrete Lehmann representation. Loop and Vec denote loop-based and vectorized implementations, respectively.}
\label{tab:performance_5orb}
\begin{tabular}{llcccc}
\hline\hline
& & $G$ & $P$ & $W$ & $\Sigma$ \\
& MF$+$Loop & 588.65\,s & 5938.45\,s & 5610.29\,s & 1771.45\,s \\
& MF$+$Vec & 235.79\,s &  284.46\,s &  376.77\,s &   59.79\,s \\
& DLR$+$Loop &  80.11\,s &  181.61\,s &  108.36\,s &  215.06\,s \\
& DLR$+$Vec &   2.21\,s &  137.55\,s &   65.48\,s &   27.32\,s \\
\hline\hline
\end{tabular}
\end{table}

To assess performance, we compare four distinct implementations derived from two choices of imaginary-frequency representation---the conventional Matsubara frequency grid (\textbf{MF}) and the discrete Lehmann representation (\textbf{DLR})---and two execution strategies: explicit Python for-loops (\textbf{Loop}) and fully vectorized array operations (\textbf{Vec}).

The implementation details highlight the different approaches to acceleration. The MF-based configurations (MF$+$Loop and MF$+$Vec) utilize a hybrid Fortran--Python architecture; in this setup, the Fourier transforms and the Dyson equation inversion are delegated to compiled Fortran routines, while the computationally dominant kernels---including the polarizability $P = GG$, the self-energy convolution $\Sigma = GW$, and the orbital- and momentum-index mappings---remain in the unaccelerated Python layer. In contrast, the DLR representation gains its advantage through frequency-axis compression, which significantly reduces the number of nodes required to represent propagators within a prescribed tolerance. The vectorization strategy (Vec) further optimizes the code by replacing interpreted Python loops over $\mathbf{k}$-points, orbitals, and frequency indices with batched array operations dispatched directly to optimized BLAS/LAPACK kernels via NumPy.

As shown in Table~\ref{tab:performance_5orb}, the pure-Python DLR$+$Vec configuration shows the best performance, achieving an overall speedup of approximately $60\times$ over the MF$+$Loop baseline. It is important to note that DLR$+$Vec achieves this without any recourse to compiled extensions, even in the Fourier and Dyson substeps where the MF baseline is specifically Fortran-accelerated. 

The five-orbital case amplifies the performance differences observed in the graphene benchmark, reflecting the steep orbital-index scaling of the GW kernels. The total time is reduced from 13908.85\,s (MF$+$Loop) to 232.56\,s (DLR$+$Vec), a speedup of approximately $60\times$. The Green's function evaluation again shows the most dramatic improvement, dropping from 588.65\,s to 2.21\,s ($266\times$), as the batched matrix inversion over all $\mathbf{k}$-points and DLR nodes eliminates the dominant cost of repeated small-matrix operations inside a Python loop. The screened interaction $W$ and the self-energy $\Sigma$ likewise benefit substantially, achieving speedups of $86\times$ and $65\times$, respectively.

The polarizability $P$, however, exhibits a qualitatively different behavior. While vectorization alone reduces the MF$+$Loop time dramatically (MF$+$Loop to MF$+$Vec: $21\times$), the DLR representation provides comparatively modest additional gains for $P$ in the vectorized case (DLR$+$Vec: 137.55\,s versus MF$+$Vec: 284.46\,s, a further $2\times$). This is because the polarizability $P = GG$ involves a convolution over fermionic orbital indices whose cost scales as $\mathcal{O}(N_\mathrm{orb}^4)$, and this orbital-index contraction remains the dominant cost regardless of the frequency-axis compression provided by DLR. 
As a result, $P$ constitutes the primary computational bottleneck in the five-orbital regime under DLR$+$Vec, in contrast to the graphene case where the two-orbital structure rendered $P$ subdominant.
This observation identifies the polarizability evaluation as the natural target for further optimization in multi-orbital systems, for example through $\mathbf{k}$-point parallelization or low-rank approximations to the orbital contraction.

\section{Discussion}\label{sec3}

We present QAssemble, a pure-Python framework for the free-energy 
functional approach to the quantum many-body problem that implements 
tight-binding, Hartree--Fock, and $GW$ approximations within a unified 
object-oriented architecture. Each physical concept---crystal structure, Hamiltonian, 
Green's function, self-energy, polarizability, and screened Coulomb 
interaction---is encapsulated as a distinct class with a well-defined 
interface, yielding a transparent correspondence between the formal 
objects of many-body theory and the numerical kernels that realize them. 
Validation on graphene with local and non-local interactions reproduces 
the expected progression from mean-field renormalization to dynamical 
correlation effects including lifetime broadening of spectral features, 
confirming the correctness of the implementation across the full 
tight-binding, Hartree--Fock, and $GW$ hierarchy.

The central technical claim of this work is that pure-Python implementation 
does not preclude practical efficiency for free-energy functional 
calculations. Performance benchmarks on both the two-orbital graphene model and a five-orbital system demonstrate the effectiveness of our approach. Systematic vectorization of the performance-critical kernels---specifically the polarizability bubble, Dyson inversion, and self-energy convolution---is combined with the discrete Lehmann representation. Together, these optimizations achieve speedups of up to $60\times$ relative to a loop-based Matsubara implementation. Once the frequency axis is 
compressed to a handful of DLR nodes and the dominant kernels are recast 
as batched tensor contractions dispatched to BLAS/LAPACK through NumPy, 
the remaining Python-level overhead becomes subdominant for the model 
Hamiltonian systems considered here. This establishes that the perceived 
dichotomy between transparent scripting-language implementations and 
compiled production codes is not fundamental, at least within the regime 
accessible to current vectorized numerical libraries.

By prioritizing code clarity, modular class hierarchies, and standardized 
HDF5 persistence, QAssemble is positioned to support both method 
development and production calculations for correlated quantum materials, 
and to accommodate future extensions toward DMFT, EDMFT, $GW$+EDMFT, and more on the same architectural foundation.

\section{Methods}\label{sec4}
The full many-body Hamiltonian is given by:
\begin{equation}
    \begin{split}
        &H_{ij\sigma}^{p, q}\left(\mathbf{R}-\mathbf{R}^\prime\right) = 
        -\sum_{\mathbf{R}\mathbf{R}^\prime}\sum_{p, q}\sum_{ij\sigma} t^{pq}_{ij}\left(\mathbf{R}-\mathbf{R}^\prime\right) {c^{\dagger}}^{p}_{i\sigma}\left(\mathbf{R}\right) {c}^{q}_{j\sigma}\left(\mathbf{R}^\prime\right) \\
        &\quad +\frac{1}{2}\sum_{\mathbf{R}\mathbf{R}^\prime}\sum_{p, q}\sum_{ijkl}\sum_{\sigma \sigma^\prime} V_{ijkl}^{p \sigma ; q \sigma^\prime} \left(\mathbf{R}-\mathbf{R}^\prime\right){c^\dagger}^{p}_{i \sigma}\left(\mathbf{R}\right) {c^\dagger}^{q}_{j \sigma^\prime}\left(\mathbf{R}^\prime\right) {c}^{q}_{k\sigma^\prime}\left(\mathbf{R}^\prime\right)c^{p}_{l\sigma}\left(\mathbf{R}\right)
    \end{split}
    \label{eq:many-body-hamiltonian}
  \end{equation}
For the definition of each symbols, please see Table \ref{tab:index-notation}. The first term constitutes the non-interacting part $H_{0}$, while the last term represents electron-electron interactions $H_{int}$. Here, $t^{p q}_{ij}\left(\mathbf{R}-\mathbf{R}^\prime\right)$ represents hopping amplitudes between orbital $(\mathbf{R},p, i)$ and $(\mathbf{R}^\prime,q, j)$. $V_{ijkl}^{p \sigma ; q \sigma^\prime}$ are the matrix elements of the two-body interaction. 

We implement electron-electron interactions in both local and non-local forms. For local interactions, we support the Slater\cite{slater_quantum_1960} and Kanamori\cite{kanamori_electron_1963} parametrizations, along with transformations between the two\cite{van_der_marel_electron-electron_1988,strand_valence-skipping_2014}. Non-local interactions can be specified through two approaches. In the user-defined approach, interactions between specific orbital pairs are needed. Alternatively, the model-based approach generates interactions using either the Ohno\cite{ohno_remarks_1964} or JTH\cite{hay_orbital_1975} potentials, where the JTH potential adopts the same functional form as Ohno but allows the same onsite Coulomb interaction value to be assigned consistently to all equivalent orbitals.

\begin{table}[!h]
    \caption{List of symbols.}
    \label{tab:index-notation}
        \begin{tabular}{ll}
            \hline
            $\mathbf{R}$& Lattice vector\\
            $\mathbf{k}, \mathbf{q}$ & Crystal momentum vector\\
            $\omega_n$& Fermionic Matsubara frequency\\
            $\nu_n$ & Bosonic Matsubara frequency\\
            $\tau$ & Imaginary time\\
            $p, q, \ldots$ & Basis site index\\
            $i,j, \ldots$ & Orbital index.\\
            $\tau_p, \tau_q, \ldots$ & Basis vector\\            
            $\sigma$ & Spin index \\
            \hline
        \end{tabular}
\end{table}
\vspace{-5pt}

For periodic systems, transformations between momentum space $\mathbf{k}$ and real space $\mathbf{R}$ are performed via discrete Fourier transforms:
\begin{equation}
    \begin{split}
        &H^{pq}_{ij\sigma}(\mathbf{k}) = \sum_{\mathbf{R}}H^{pq}_{ij\sigma}(\mathbf{R})e^{-i\mathbf{k} \cdot \left(\mathbf{R} + \tau_{p} - \tau_{q}\right)}\\
        &H^{pq}_{ij\sigma}(\mathbf{R}) = \frac{1}{N_{\mathbf{k}}}\sum_{\mathbf{k}}H^{pq}_{ij\sigma}(\mathbf{k})e^{i\mathbf{k} \cdot \left(\mathbf{R} +\tau_{p} - \tau_{q}\right)}
    \end{split}
    \label{eq:fourier-transform-kr}
\end{equation}

where $N_{\mathbf{k}}$ is the number of $\mathbf{k}$ points. Similar transforms apply to Green's functions, self-energies, and interaction matrices. These transformations are efficiently implemented using fast Fourier transform (FFT) algorithms \cite{frigo_fftw_1998}.

A central quantity in many-body theory is the single-particle Green's function. At the finite temperature $T = 1/(k_b \beta)$, we construct Green's function. The Green's function ${G_0}^{p, q}_{ij \sigma}(\tau)$ for imaginary time $0 \le \tau < \beta$ is defined as:

\begin{equation}
    \begin{split}
        &{G_0}^{pq}_{ij\sigma}(\mathbf{R} - \mathbf{R}^\prime, \tau - \tau^\prime) = - \langle T_{\tau} c^{p}_{i\sigma}\left(\mathbf{R}, \tau\right){c^{\dagger}}^{q}_{j\sigma}\left(\mathbf{R}^\prime, \tau^\prime \right) \rangle_0
    \end{split}
    \label{eq:bare-green-tau}
\end{equation}
where $T_{\tau}$ is the imaginary-time ordering operator. The expectation value, $\langle \cdots \rangle_0$,  is taken in grand-canonical ensemble of $H_0$. This Green's function describes the propagation of an electron (hole) added at orbital $(\mathbf{R}^\prime, q, j)$ ($(\mathbf{R}, p, i)$) at imaginary time $\tau^\prime$ ($\tau$) and removed at orbital $(\mathbf{R}, p, i)$ ($(\mathbf{R}^\prime, q, j)$) with imaginary time $\tau$ ($\tau^\prime$).
From real-space bare Green's function in imaginary time, we can compute $\mathbf{k}$-space bare Green's function in Matsubara frequency. 
\begin{equation}
    \begin{split}
        &{G_0}^{pq}_{ij\sigma}(\mathbf{k}, i\omega_n) = \\
        &\qquad \qquad \frac{1}{N_\mathbf{k}} \int_{0}^{\beta} d(\tau - \tau^\prime) \sum_{\mathbf{R}, \mathbf{R}^\prime}{G_0}^{pq}_{ij\sigma}(\mathbf{R} - \mathbf{R}^\prime, \tau - \tau^\prime)e^{-i(\mathbf{k} \cdot (\mathbf{R}-\mathbf{R}^\prime) + \omega_n \tau)} 
    \end{split}
    \label{eq:bare-green-freq}
\end{equation}
We used discrete Lehmann representation (DLR)\cite{kaye_discrete_2022} for the compact representation of imaginary-time and Matsubara frequency Green's function. 
The interacting Green's function $G$ is obtained through the Dyson equation :
\begin{equation}
    \begin{split}
        &G^{pq}_{ij\sigma}(\mathbf{k}, i\omega_n) = \\
        &\qquad \qquad {G_0}^{pq}_{ij\sigma}(\mathbf{k}, i\omega_n) + \sum_{r, s} \sum_{k, l}
        {G_0}^{pr}_{ik\sigma}(\mathbf{k}, i\omega_n)
        \Sigma^{rs}_{kl\sigma}(\mathbf{k}, i\omega_n)
        G^{sq}_{lj\sigma}(\mathbf{k}, i\omega_n)
    \end{split}
    \label{full-green-freq}
\end{equation}
where $\Sigma$ is the irreducible electron self-energy. 

\subsection{Hartree-Fock}
The Hartree-Fock (HF) method\cite{hartree_self-consistent_1997,slater_self_1928,slater_note_1930,slater_simplification_1951,froese_fischer_general_1987} is a mean-field approximation that treats electron-electron interactions through static effective potentials. 
The Hartree term represents the classical electrostatic potential felt by an electron due to the charge density of other electrons. The Hartree self-energy is :
\begin{equation}
    \begin{split}
        &{\Sigma^H}_{ij\sigma}^{pr} (\mathbf{k}) = \sum_{\sigma^\prime}\sum_{q} \sum_{kl} V_{ijkl}^{p\sigma ; q\sigma^\prime}(\mathbf{k} = 0) \\
        &\qquad \qquad \times \frac{1}{N_{\mathbf{k}}}\sum_{\mathbf{k}^\prime} n^{q}_{lk\sigma^\prime}(\mathbf{k}^\prime)\delta_{pr}
    \end{split}
    \label{eq-hartree-self-energy}
  \end{equation}
  
where $N_{\mathbf{k}}$ is the number of $\mathbf{k}$-points in the Brillouin zone, $n^{q}_{lk\sigma^\prime}(\mathbf{k}^\prime)$ represents the momentum-resolved density, and the interaction $V_{ijkl}^{p\sigma ; q\sigma^\prime}(\mathbf{k} = 0)$ is evaluated at zero momentum transfer. 

The Fock term accounts for exchange interactions arising from the antisymmetry of the fermionic wavefunction, providing a non-local correction. The Fock self-energy is :
\begin{equation}
    \begin{split}
        &{\Sigma^F}^{p;q}_{ij\sigma}(\mathbf{k}) = - \sum_{\mathbf{R}}\sum_{kl}n^{q;p}_{lk\sigma}(\mathbf{R}) \\
        &\qquad \qquad \times V_{ijkl}^{p\sigma;q\sigma^\prime}(\mathbf{R}) \delta_{\sigma\sigma^\prime} e^{i \mathbf{k} \cdot \mathbf{R}}
    \end{split}
    \label{eq-fock-self-energy}
\end{equation}
where $n^{q;p}_{lk\sigma}(\mathbf{R})$ is the density matrix in real space connecting sites separated by lattice vector $\mathbf{R}$, and $\delta_{\sigma\sigma^\prime}$ ensures that exchange occurs only between electrons of the same spin. This term is generally non-diagonal in both real space and orbital indices, and is responsible for phenomena like exchange splitting and magnetic ordering.

The total HF self-energy is $\Sigma^{HF} = \Sigma^{H} + \Sigma^{F}$. Then the HF Hamiltonian is :
\begin{equation}
    \begin{split}
        &H_{HF}(\mathbf{k}) = H_0(\mathbf{k}) + \Sigma^{HF}(\mathbf{k}) - \mu \hat{I}
    \end{split}
    \label{hf-hamiltonian}
\end{equation}
  where $\mu$ is the chemical potential of the system
The diagrammatic representations of the Hartree and Fock self-energies are shown in Extended Data Fig.~\ref{fig-hartree-fock-self}.



\subsection{$GW$ approximation}
$GW$ approximation\cite{hedin_new_1965,hybertsen_electron_1986,godby_self-energy_1988,massidda_band-structure_1995,massidda_quasiparticle_1997,aulbur_quasiparticle_2000,fulde_semiconductors_1995} goes beyond Hartree-Fock by including frequency-dependent (dynamical) screening of the Coulomb interaction. The name "$GW$" reflects that the self-energy is constructed as a convolution of the Green's function $G$ and the dynamically screened interaction $W$.

For given $G$, we compute several quantities, such as irreducible polarizability, screened Coulomb interaction, and the irreducible self-energy. The irreducible polarizability $P$ in the Matsubara frequency domain is:
\begin{equation}
    \begin{split}
        &P^{p\sigma;q\sigma^\prime}_{ijkl}(\mathbf{k}, i\nu_n) = \int_{0}^{\beta} d\tau \sum_{\mathbf{R}} G_{ki\sigma^\prime}^{qp}(-\mathbf{R}, -\tau) \\
        &\qquad \qquad \times G^{pq}_{lj\sigma}(\mathbf{R},\tau) \delta_{\sigma \sigma^\prime} e^{i(\mathbf{k} \cdot \mathbf{R} - \nu_n \tau)}
    \end{split}
    \label{eq-polarizability}
\end{equation}
where $i\nu_n = 2n\pi/\beta$ are bosonic Matsubara frequencies, and the integration over imaginary time $\tau$ perform the convolution of two fermionic Green's functions. The Kronecker delta $\delta_{\sigma \sigma^\prime}$ indicates that the electron and hole must have same spin in the non-interacting bubble.

The screened interaction $W$ accounts for how the bare Coulomb interaction $V$ is reduced (screened) by the polarization of the surrounding electron gas. The screened interaction is obtained by the Dyson's equation :
\begin{equation}
    \begin{split}
        &W^{p\sigma ; q\sigma^\prime}_{ijkl}(\mathbf{k}, i\nu_n) = V^{p\sigma ; q\sigma^\prime}_{ijkl}(\mathbf{k}) \\
        &\qquad + \sum_{rs} \sum_{i^\prime j^\prime k^\prime l^\prime}V^{p\sigma ; r\sigma^\prime}_{ii^\prime j^\prime l}(\mathbf{k})P^{r\sigma ; s\sigma^\prime}_{i^\prime k^\prime l^\prime j^\prime}(\mathbf{k}, i\nu_n)W^{s\sigma ; q\sigma^\prime}_{k^\prime jkl^\prime}(\mathbf{k}, i\nu_n)
    \end{split}
    \label{eq-screened-coulomb}
\end{equation}
The matrix equation must be solved for each momentum $\mathbf{k}$ and bosonic frequency $i\nu_n$ to obtain the full frequency-dependent screened interaction.

The $GW$ self-energy incorporates dynamic correlations through the frequency-dependent screened interaction. The correlation part of the $GW$ self-energy is given by:
\begin{equation}
    \begin{split}
        &{\Sigma^{C, GW}}_{ij\sigma}^{pq}(\mathbf{k}, i\omega_n) = -\int_{0}^{\beta}d\tau \sum_{\mathbf{R}}\sum_{kl}G^{qp}_{lk\sigma}(\mathbf{R}, \tau) \\
        &\qquad \qquad \times {W^C}^{p\sigma q\sigma^\prime}_{ijkl}(\mathbf{R},\tau)\delta_{\sigma\sigma^\prime} e^{i(\mathbf{k}\cdot\mathbf{R} - \omega_n \tau)}
    \end{split}
    \label{eq-gw-self-energy}
\end{equation}
where $W^{C} = W - V$ is the dynamical part of the screened interaction.

The diagrammatic representation of the $GW$ approximation is shown in Extended Data Fig.~\ref{fig-GW-Diag}.

\section*{Code availability}
The QAssemble source code is openly available at \url{https://www.qassemble.org}, 
where comprehensive documentation, installation instructions, tutorials, and 
usage examples covering all methods presented in this work are provided.

\section*{Data availability}
The input files and parameter 
settings required to reproduce these results are distributed as part of 
the example suite accompanying the QAssemble source code.

\section{Acknowledgments}
For the code development and validation, we used resources of the Center for Advanced Computation at Korea Institute for Advanced Study and the National
Energy Research Scientific Computing Center (NERSC), a
U.S. Department of Energy Office of Science User Facility
operated under Contract No. DE-AC02-05CH11231. SC was
supported by a KIAS Individual Grant (CG090601) at Korea
Institute for Advanced Study. M.H. is supported by a KIAS
Individual Grant (No. CG091301) at Korea Institute for Advanced Study. SC and SM were supported by Quantum Simulator Development
Project for Materials Innovation through the National Research Foundation of Korea (NRF) funded by the Korean government (Ministry of Science and ICT(MSIT))(No. NRF-2023M3K5A1094813).

\bibliography{ref}

\clearpage

\setcounter{figure}{0}
\renewcommand{\figurename}{Extended Data Fig.}
\renewcommand{\thefigure}{\arabic{figure}}

\begin{figure*}
  \centering
  \includegraphics[width=0.7\columnwidth]{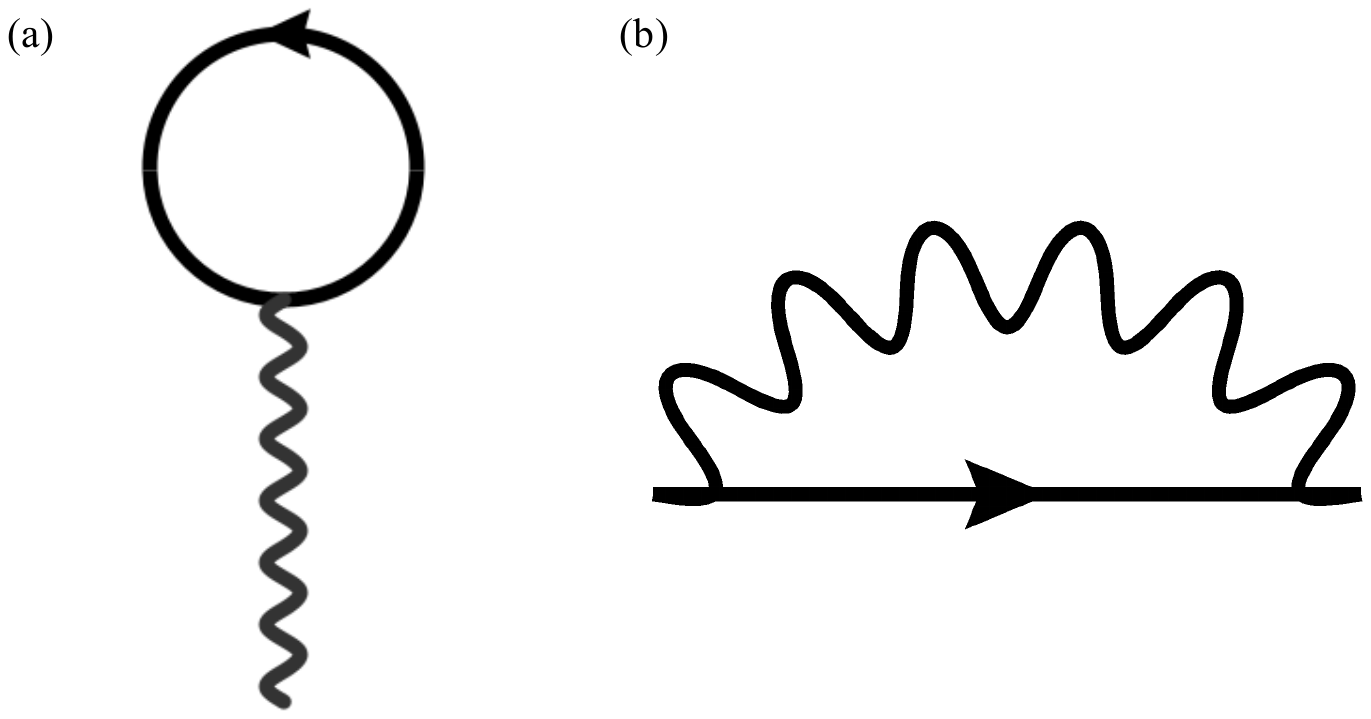}
  \caption{\textbf{Diagrammatic representation of the Hartree--Fock 
  self-energies.} (a) The Hartree self-energy is depicted as a single 
  fermion line (the propagator) with a wavy interaction line forming a 
  closed loop back to the same propagator, representing the interaction 
  with the average density. (b) The Fock self-energy corresponds to an 
  exchange diagram.}
  \label{fig-hartree-fock-self}
\end{figure*}

\begin{figure*}
  \centering
  \includegraphics[width=0.7\columnwidth]{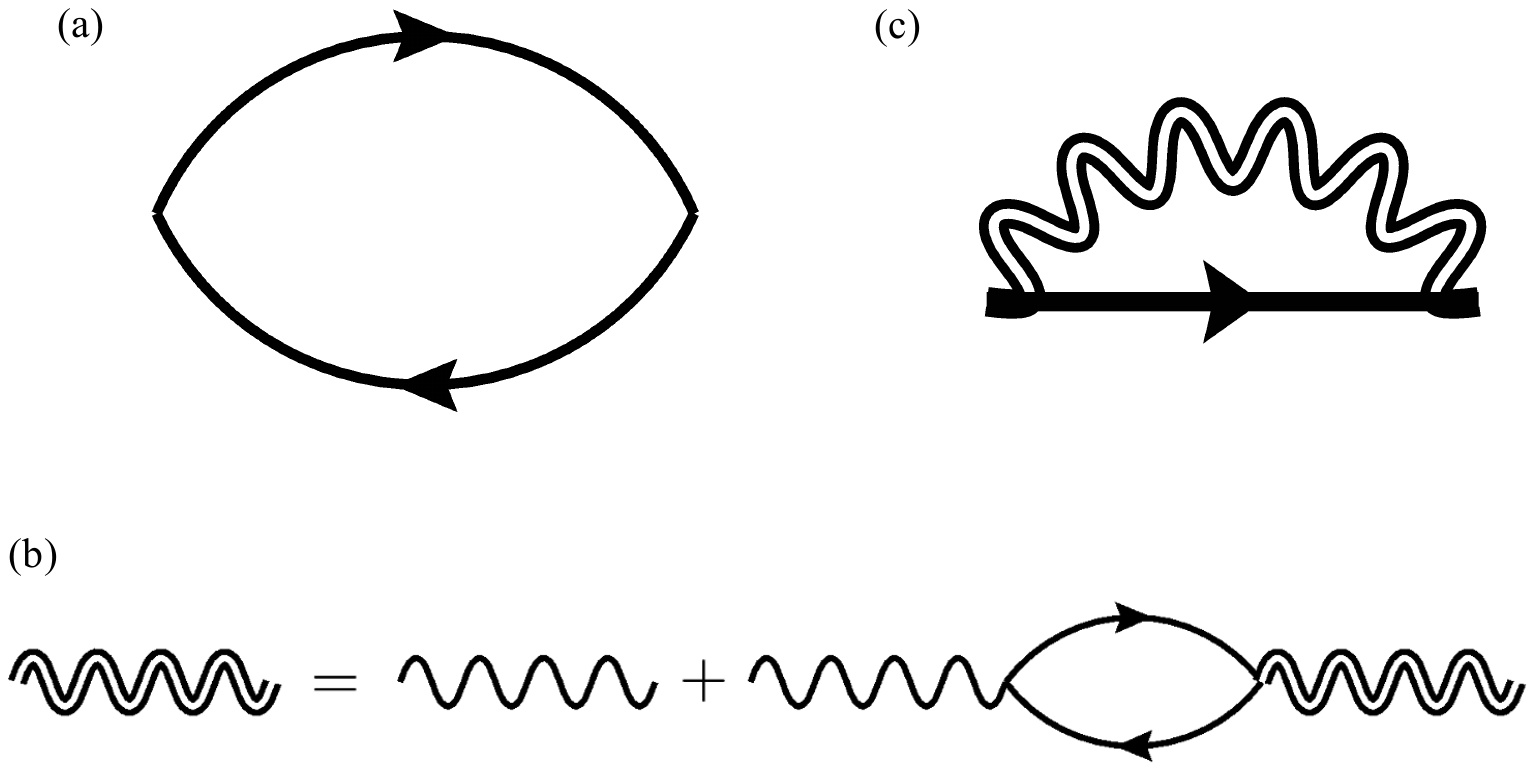}
  \caption{\textbf{Diagrammatic representation of the $GW$ approximation.}
  (a) Polarizability $P$, approximated by a single particle-hole bubble 
  (loop of two Green's functions with opposite directions). (b) Dyson 
  equation for the screened Coulomb interaction $W$, given by 
  $W = V + VPW$, where $V$ is the bare Coulomb interaction (wavy line) 
  and $P$ is the irreducible polarizability from (a). (c) $GW$ self-energy 
  $\Sigma = -GW$, represented as the product of one Green's function 
  line $G$ and one screened interaction line $W$.}
  \label{fig-GW-Diag}
\end{figure*}

\end{document}


\title[Article Title]{Supplementary Information for "\texttt{QAssemble:} A Pure Python Package for Quantum Many-Body Theory"}

\author[1,2]{\fnm{Seongjun} \sur{Mo}}

\author[3]{\fnm{Dongming} \sur{Li}}

\author[2]{\fnm{Mancheon} \sur{Han}}

\author[2,4]{\fnm{Johan} \sur{Jönsson}}

\author[5,6]{\fnm{Byungkyun} \sur{Kang}}

\author[1]{\fnm{Hoonkyung} \sur{Lee}}

\author[5,7]{\fnm{Gabriel} \sur{Kotliar}}

\author*[2]{\fnm{Sangkook} \sur{Choi}}

\affil[1]{\orgdiv{Advanced Materials Program, Department of Physics}, \orgname{Konkuk University}, \orgaddress{\city{Seoul}, \postcode{05029}, \country{Korea}}}

\affil[2]{\orgdiv{School of Computational Sciences}, \orgname{Korea Institute for Advanced Study}, \orgaddress{\city{Seoul}, \postcode{02455}, \country{Korea}}}

\affil[3]{\orgdiv{Department of Electrical and Computer Engineering}, \orgname{University of Massachusetts Amherst}, \orgaddress{\street{300 Massachusetts Ave}, \city{Amherst}, \state{MA}, \postcode{01003}, \country{USA}}}

\affil[4]{\orgdiv{Institute of Physics, Faculty of Physics, Astronomy and Informatics}, \orgname{Nicolaus Copernicus University}, \orgaddress{\city{Toruń}, \postcode{87-100}, \country{Poland}}}

\affil[5]{\orgdiv{Condensed Matter Physics and Materials Science Department}, \orgname{Brookhaven National Laboratory}, \orgaddress{Upton, NY} \postcode{11973}, \country{USA}}

\affil[6]{\orgdiv{College of Arts and Sciences}, \orgname{University of Delaware}, \orgaddress{Newark, DE}, \postcode{19716}, \country{USA}}

\affil[7]{\orgdiv{Department of Physics and Astronomy}, \orgname{Rutgers University}, \orgaddress{NJ}, \postcode{08854}, \country{USA}}

\maketitle

\section{Database architecture}

\begin{figure*}[!htb]
  \centering
    \includegraphics[width=1.0\columnwidth]{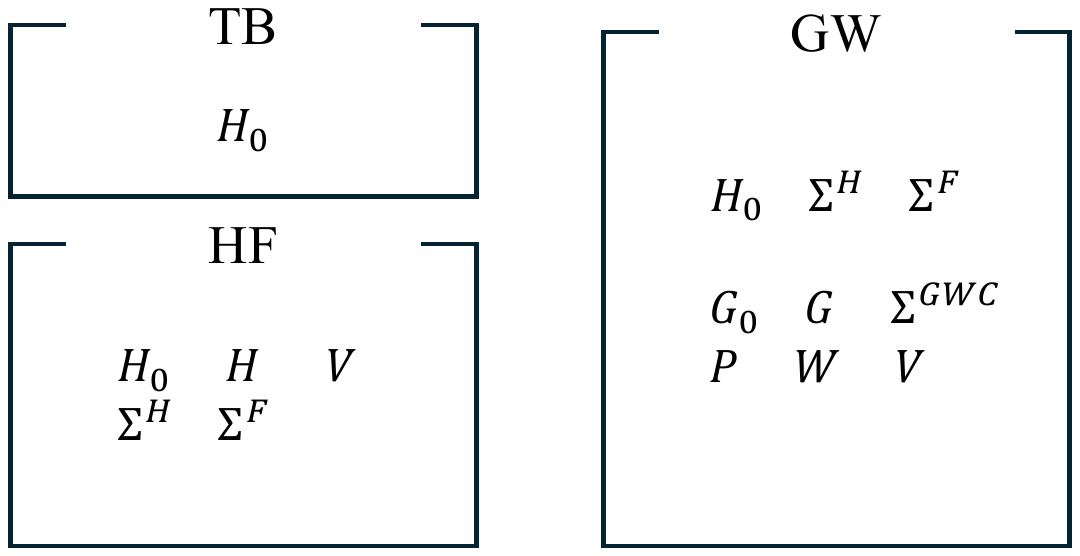}

  \caption{\textbf{HDF5 data architecture of QAssemble.} Grouped data structure about QAssemble.}
  \label{fig-data-structure}
\end{figure*}
QAssemble supports writing and reading data with HDF5 IO library. Each component stores data in a single HDF5 file called "glob.h5" by default. However, a user can replace the file name. Keys are provided to ensure the data is easily retrieved when needed. There are four HDF5 groups included in the file : "input", "tb", "hf", and "gw". Data in each method is stored in its corresponding HDF5 group. Figure \ref{fig-data-structure} provides an overview of the groups and important data stored in "glob.h5".

\section{Performance Benchmark: Graphene Extended Hubbard Model}

Table~\ref{tab:performance} reports the wall-clock time per GW iteration for the four principal quantities---the interacting Green's function $G$, the irreducible polarizability $P$, the screened Coulomb interaction $W$, and the correlation self-energy $\Sigma$---evaluated on the graphene benchmark of the \nameref{sec:validation} section. The four computational configurations (MF$+$Loop, MF$+$Vec, DLR$+$Loop, DLR$+$Vec) are those introduced in the main text, combining the conventional Matsubara frequency grid (MF) or the discrete Lehmann representation (DLR) with loop-based (For) or vectorized (Vec) implementations. We refer the reader to the main text for a detailed discussion of the hybrid Fortran--Python architecture of the MF-based configurations and the pure-Python nature of DLR$+$Vec; here we focus on how these strategies perform in the minimal two-orbital setting, which serves as a complementary reference point to the five-orbital benchmark in the main text.

\begin{table}[h]
\centering
\caption{Wall-clock time (seconds) per GW iteration for each physical quantity, measured on the graphene benchmark ($25\times25\times1$ $k$-grid, $\Lambda=100$\,eV). MF denotes the conventional Matsubara frequency grid with Fortran-based routines; DLR denotes the discrete Lehmann representation. Loop and Vec denote loop-based and vectorized implementations, respectively.}
\label{tab:performance}
\begin{tabular}{lcccc}
\hline\hline
& $G$ & $P$ & $W$ & $\Sigma$ \\
\hline
MF$+$Loop  & 36.13\,s & 12.35\,s & 12.38\,s & 10.97\,s \\
MF$+$Vec  & 17.96\,s &  1.54\,s &  6.49\,s &  3.40\,s \\
DLR$+$Loop & 42.82\,s &  2.55\,s &  3.91\,s &  2.00\,s \\
DLR$+$Vec &  0.43\,s &  2.02\,s &  0.64\,s &  0.04\,s \\
\hline\hline
\end{tabular}
\end{table}

The results in Table~\ref{tab:performance} show that the combination of DLR and vectorization (DLR$+$Vec) achieves the best overall performance, reducing the total iteration time from 71.83\,s (MF$+$Loop) to 3.14\,s---a speedup of approximately $23\times$. Consistent with the five-orbital case, this pure-Python implementation outperforms the Fortran-assisted MF$+$Loop baseline across all four quantities. The most pronounced gains appear in $G$ and $\Sigma$, where the Dyson equation inversion benefits simultaneously from the compressed DLR frequency axis and the batched matrix inversion over all $\mathbf{k}$-points and frequency nodes, reducing the evaluation times from 36.13\,s to 0.43\,s ($84\times$) and from 10.97\,s to 0.04\,s ($274\times$), respectively. For the polarizability $P$, vectorization alone already yields a substantial reduction (MF$+$Loop to MF$+$Vec: $8\times$), and the DLR representation provides a further reduction in the loop-based case (MF$+$Loop to DLR$+$Loop: $4.8\times$), though the two effects do not combine multiplicatively for $P$ due to the differing frequency-axis structure of the bosonic bubble.

A notable exception is the Green's function under DLR$+$Loop (42.82\,s), which is slower than MF$+$Loop (36.13\,s). Although the DLR basis compresses the imaginary-frequency axis significantly, this reduction in the number of frequency points is entirely offset by the Python for-loop overhead when the Dyson equation inversion is not vectorized, exposing the cost of repeated small matrix operations inside an interpreted loop. This makes explicit that DLR and vectorization are complementary optimizations that must be applied together to realize their combined benefit: neither the frequency-axis compression of DLR nor the batched array dispatch of NumPy is individually sufficient to overcome the Fortran-assisted MF$+$Loop baseline for $G$, but their combination yields nearly two orders of magnitude of improvement. Compared to the five-orbital case discussed in the main text, the graphene benchmark represents a regime in which $P$ is subdominant---a direct consequence of the minimal two-orbital structure, which suppresses the $\mathcal{O}(N_\mathrm{orb}^4)$ orbital contraction cost that dominates the multi-orbital regime.

These results confirm that the vectorization strategy described in the \nameref{sec:vectorization} section achieves practical efficiency within a pure-Python environment, even relative to implementations that rely on compiled Fortran kernels.